\documentclass[11pt]{article}
\usepackage[T1]{fontenc}

\usepackage[latin1]{inputenc}

\usepackage[english,french]{babel}
\usepackage{color}
\usepackage{graphicx}

\usepackage{dcolumn}

\usepackage{moreverb}

\usepackage{amsmath,amssymb,amsfonts}

\begin{document}
\numberwithin{equation}{section}
\newcommand{\boxedeqn}[1]{%
  \[\fbox{%
      \addtolength{\linewidth}{-2\fboxsep}%
      \addtolength{\linewidth}{-2\fboxrule}%
      \begin{minipage}{\linewidth}%
      \begin{equation}#1\end{equation}%
      \end{minipage}%
    }\]%
}


\newsavebox{\fmbox}
\newenvironment{fmpage}[1]
     {\begin{lrbox}{\fmbox}\begin{minipage}{#1}}
     {\end{minipage}\end{lrbox}\fbox{\usebox{\fmbox}}}

\begin{flushleft}

\title*{\textbf{Polynomial Associative Algebras for Quantum Superintegrable Systems with a Third Order Integral of
Motion}}
\newline
\newline
Ian Marquette
\newline
D\'epartement de physique et Centre de recherche math\'ematique,
Universit\'e de Montr\'eal,
\newline
C.P.6128, Succursale Centre-Ville, Montr\'eal, Qu\'ebec H3C 3J7,
Canada
\newline
ian.marquette@umontreal.ca
\newline
\newline
\newline
We consider a superintegrable Hamiltonian system in a
two-dimensional space with a scalar potential that allows one
quadratic and one cubic integral of motion. We construct the most
general associative cubic algebra and we present specific
realizations. We use them to calculate the energy spectrum. All
classical and quantum superintegrable potentials separable in
cartesian coordinates with a third order integral are known. The
general formalism is applied to one of the quantum potentials.
\newline
\newline
\section{Introduction}
The purpose of this article is to study the algebra of integrals of
motion of a certain class of quantum superintegrable systems
allowing a second and a third order integral of motion. We will
consider a cubic associative algebra and we will study its algebraic
realization. A systematic search for superintegrable systems in
classical and quantum mechanics was started some time
ago$^{8,10,18}$. The study of superintegrable system with a third
order integral is more recent. All classical and quantum
superintegrable potentials in E(2) that separate in cartesian
coordinate and allow a third order integral were found by
S.Gravel$^{13}$. In this article we will be interested in particular
in one new potentials found in Ref [13] that was studied by the
dressing chain method in Ref [14].
\newline
It is well known that in quantum mechanics the operators commuting
with the Hamiltonian, form an o(4) algebra for the hydrogen
atom$^{1}$ and a u(3) algebra for the harmonic oscillator$^{15}$.
This type of symmetry is called a dynamical or hidden symmetry and
can be used give a complete description of the quantum mechanical
system. The symmetry determines all the quantum numbers, the
degeneracy of the energy levels and the energy spectrum$^{9,15}$.
\newline
Many examples of polynomial algebras have been used in different
branch of physics. C.Daskaloyannis studied the quadratic Poisson
algebras of two-dimensional classical superintegrable systems and
quadratic associative algebras of quantum superintegrable
systems$^{2,3,4}$. He shows how the quadratic associative algebras
provide a method to obtain the energy spectrum. He uses realizations
in terms of a deformed oscillator algebra$^{5}$. We will follow an
analogous approach for the study of cases with third order
integrals.
\newline
In an earlier article$^{17}$ we considered cubic Poisson algebras
for classical potentials and applied the theory to the 8 potentials
separating in cartesian coordinates and allowing a third order
integral. The purpose of this article is to study cubic associative
algebras. We find the realization of these polynomial associative
algebras in terms of a parafermionic algebra. From this we find Fock
type representations and the energy spectrum. We reduce this problem
to the problem of solving two algebraic equations. This article
provides another example of the fact that it is very useful to
consider not only Lie algebras in the study of quantum systems but
also polynomial algebras.
\section{Cubic Associative algebras and their algebraic
realizations} We consider a quantum superintegrable system with a
quadratic Hamiltonian and one second order and one third order
integral of motion.
\newline
\newline
\begin{equation*}
H = a(q_{1},q_{2})P_{1}^{2} + 2b(q_{1},q_{2})P_{1}P_{2} +
c(q_{1},q_{2})P_{2}^{2} + V(q_{1},q_{2})
\end{equation*}
\begin{equation*}
A = A(q_{1},q_{2},P_{1},P_{2})= d(q_{1},q_{2})P_{1}^{2} +
2e(q_{1},q_{2})P_{1}P_{2} +
\end{equation*}
\begin{equation}
f(q_{1},q_{2})P_{2}^{2} + g(q_{1},q_{2})P_{1} + h(q_{1},q_{2})P_{2}
+ Q(q_{1},q_{2})
\end{equation}
\begin{equation*}
B = B(q_{1},q_{2},P_{1},P_{2}) = u(q_{1},q_{2})P_{1}^{3} +
3v(q_{1},q_{2})P_{1}^{2}P_{2} + 3w(q_{1},q_{2})P_{1}P_{2}^{2}
\end{equation*}
\begin{equation*}
+x(q_{1},q_{2})P_{2}^{3}+j(q_{1},q_{2})P_{1}^{2}+
2k(q_{1},q_{2})P_{1}P_{2} + l(q_{1},q_{2})P_{2}^{2} +
\end{equation*}
\begin{equation*}
m(q_{1},q_{2})P_{1} + n(q_{1},q_{2})P_{2} + S(q_{1},q_{2})
\end{equation*}
with
\newline
\begin{equation}
P_{1}=-i\hbar \partial_{1} , P_{2}=-i\hbar \partial_{2}
\end{equation}
\begin{equation}
[H,A]=[H,B]=0
\end{equation}
\newline
We assume that our integrals close in a polynomial algebra:
\newline
\[ [A,B]=C \]
\begin{equation}
[A,C] = \alpha A^{2} + \beta \{A,B\} + \gamma A + \delta B +
\epsilon
\end{equation}
\[ [B,C] = \mu A^{3} + \nu A^{2} + \rho B^{2} + \sigma \{A,B\} + \xi A + \eta B + \zeta \]
\newline
where $\{\}$ denotes an anticommutator.
\newline
\newline
The Jacobi identity $[A,[B,C]]=[B,[A,C]] $ implies $ \rho = -\beta$
, $\sigma= - \alpha$ and $\eta = - \gamma $.
\newline
\[ [A,B]=C \]
\begin{equation}
 [A,C]=\alpha A^{2} + \beta \{A,B\} + \gamma A + \delta B + \epsilon
\end{equation}
\[ [B,C]=\mu A^{3} + \nu A^{2} - \beta B^{2} - \alpha \{A,B\} + \xi A -
\gamma B + \zeta \quad . \]
\newline
\newline
The Casimir operator of a polynomial algebra is an operator that
commutes with all elements of the associative algebra. The Casimir
operator satisfies:
\begin{equation}
[K,A]=[K,B]=[K,C]=0
\end{equation}
\newline
and this implies
\newline
\[ K = C^{2} - \alpha \{A^{2},B\} - \beta \{A,B^{2}\} + (\alpha
\beta - \gamma)\{A,B\} + (\beta^{2} - \delta)B^{2} \]
\begin{equation}
(+\beta \gamma - 2\epsilon)B+\frac{\mu}{2}A^{4} +
\frac{2}{3}(\nu+\mu \beta)A^{3}+(-\frac{1}{6}\mu \beta^{2} +
\frac{\beta \nu}{3} + \frac{\delta \mu}{2} + \alpha^{2} + \xi)A^{2}
\end{equation}
\[+(-\frac{1}{6}\mu \beta \delta + \frac{\delta \nu}{3} + \alpha
\gamma + 2\zeta)A \]
\newline
\newline
We construct a realization of the cubic associative algebra by means
of the deformed oscillator technique. We use a deformed oscillator
algebra $ \{b^{t},b,N\}$ which satisfies the relation
\begin{equation}
[N,b^{t}]=b^{t} , [N,b]=-b , b^{t}b=\Phi(N) , bb^{t}=\Phi(N+1)
\end{equation}
We request that the "structure function" $\Phi(x)$ should be a real
function that satifies the boundary condition $ \Phi(0)=0$, with
$\Phi(x) > 0$ for $x > 0 $. These constraints imply the existence of
a Fock type representation of the deformed oscillator
algebra$^{4,5}$. There is a Fock basis $|n>$ , n=0,1,2... satisfying
\newline
\begin{equation}
N|n>=n|n>,\quad b^{t}|n>=\sqrt{\Phi(N+1)}|n+1>
\end{equation}
\begin{equation}
b|0>=0,\quad b|n>=\sqrt{\Phi(N)}|n-1>
\end{equation}
\newline
We consider the case of a nilpotent deformed oscillator algebra,
i.e., there should be an a integer p such that,
\newline
\begin{equation}
b^{p+1}=0 , (b^{t})^{p+1}=0
\end{equation}
These relations imply that we have
\begin{equation}
\Phi(p+1)=0
\end{equation}
In this case we have a finite-dimensional representation of
dimension p+1.
\newline
\newline
Let us show that there is a realization of the form:
\newline
\begin{equation}
A=A(N), B=b(N)+b^{t}\rho(N)+\rho(N)b
\end{equation}
\newline
The functions A(N) , b(N) et $\rho(N)$ will be determined by the
cubic associative algebra, in particular the first and second
relation. We use the commutation relation of the cubic associative
algebra to obtain
\newline
\[ [A,B]=C \]
\begin{equation}
[A,B]=b^{t}\bigtriangleup A(N) \rho(N) - \rho(N)\bigtriangleup A(N)
b
\end{equation}
\newline
\[ \bigtriangleup A(N)= A(N+1) - A(N) \]
\newline
\[ [A,C]=\alpha A^{2} + \beta \{A,B\} + \gamma A + \delta B +
\epsilon
\]
\begin{equation}
=b^{t}(\gamma(A(N+1)+A(N))+\delta)\rho(N)+
\end{equation}
\[\rho(N)(\gamma(A(N+1)+A(N))+\delta)b+\alpha A^{2}(N)+2\beta A(N)b(N)\]
\[+\gamma A(N)+\delta b(N)+\epsilon\]
\newline
using
\newline
\[ [A,C]=b^{t}\bigtriangleup A(N)\rho(N)\bigtriangleup A(N) +
\bigtriangleup A(N)\rho(N)\bigtriangleup A(N)b\]
\begin{equation}
= b^{t}\bigtriangleup A(N)^{2}\rho(N) + \rho(N)\bigtriangleup
A(N)^{2}b
\end{equation}
\newline
we obtain two equations that allow us to determine A(N) and b(N).
\newline
\begin{equation}
\bigtriangleup A(N)^{2} = \gamma(A(N+1)+A(N))+\epsilon
\end{equation}
\[\alpha A(N)^{2} + 2\gamma A(N) + b(N) + \delta A(N) + \epsilon
b(N) + \xi = 0\]
\newline
We shall distinguish two cases.
\newline
\newline
\textbf{Case 1} $ \beta \neq 0$
\begin{equation}
A(N) = \frac{\beta}{2}((N+u)^{2}- \frac{1}{4} -
\frac{\delta}{\beta^{2}})
\end{equation}
\[ b(N) = \frac{\alpha}{4}((N+u)^{2}-\frac{1}{4})+ \frac{\alpha
\delta - \gamma \beta}{2\beta^{2}}\]
\[ -\frac{\alpha \delta^{2}-2\gamma \delta \beta +
4\beta^{2}\epsilon}{4\beta^{4}}\frac{1}{(N+u)^{2}-\frac{1}{4}}\]
\newline
\newline
The constant u will be determined below using the fact that we
require that the deformed oscillator algebras should be nilpotent.
The last equation of the associative cubic algebra contains the
cubic term and is
\newline
\begin{equation}
[B,C]=\mu A^{3} + \nu A^{2} - \beta B^{2} - \alpha \{A,B\} + \xi A -
\gamma B + \zeta
\end{equation}
\newline
We obtain the equation,
\newline
\begin{equation}
2\Phi(N+1)( \bigtriangleup A(N) + \frac{\gamma}{2})\rho(N) -
2\Phi(N)(\bigtriangleup A(N-1) - \frac{\gamma}{2})\rho(N-1)
\end{equation}
\[ =  \mu A(N)^3 + \nu A(N)^2 - \beta b(N)^2 - 2\alpha A(N)b(N) + \xi A(N) -
\gamma b(N) + \zeta \]
\newline
The Casimir operator is now realized as
\newline
\begin{equation}
 K = \Phi(N+1)(\beta^{2}-\delta-2\beta A(N)-\bigtriangleup
A(N)^{2})\rho(N)
\end{equation}
\[+\Phi(N)(\beta^{2}-\delta-2\beta A(N) -
\bigtriangleup A(N-1)^{2})\rho(N-1)-2\alpha A(N)^{2}b(N) \]
\[+(\beta^{2}-\delta-2\beta A(N))b(N)^{2} + 2(\alpha \beta
-\gamma)A(N)b(N) + (\beta \gamma - 2\epsilon)b(N) +
\frac{\mu}{2}A(N)^{4}\]
\[+\frac{2}{3}(\nu+\mu \beta)A(N)^{3}+(-\frac{1}{6}\mu \beta^{2} +
\frac{\beta \nu}{3} + \frac{\delta \mu}{2} + \alpha^{2} +
\epsilon)A(N)^{2}\]
\[+(-\frac{1}{6}\mu \beta \delta + \frac{\delta a}{3} + \alpha
\gamma + 2\zeta)A(N) \]
\newline
and finally the structure function is
\newline
\[\Phi(N)=\]
\[\frac{\rho(N-1)^{-1}}{(\bigtriangleup
A(N-1)-\frac{\beta}{2})(\beta^{2}-\epsilon-2\beta
A(N)-\bigtriangleup A(N)^{2})+(\bigtriangleup
A(N)+\frac{\beta}{2})(\beta^{2}-\delta-2\beta A(N)-\bigtriangleup
A(N-1)^{2})}\]
\[((\bigtriangleup A(N)+\frac{\beta}{2})(K+2\alpha
A(N)^{2}b(N)-(\beta^{2}-\delta-2\beta A(N))b(N)^{2}\]
\begin{equation}
-2(\alpha \beta-\gamma)A(N)b(N)-(\beta \gamma
-2\epsilon)b(N)-\frac{\mu}{2}A(N)^{4}-\frac{2}{3}(\nu+\mu
\beta)A(N)^{3}
\end{equation}
\[-(-\frac{1}{6}\mu \beta^{2}+\frac{\beta
\nu}{3}+\alpha^{2}+\xi)A(N)^{2}-(-\frac{1}{6}\mu \beta \delta
+\frac{\delta \nu}{3} + \alpha \gamma + 2\zeta)A(N))\]
\[-\frac{1}{2}(\beta^{2}-\delta-2\beta A(N)-\bigtriangleup
A(N)^{2})(gA(N)^{3}+\nu A(N)^{2}-\beta b(N)^{2}-2\alpha A(N)b(N)+\xi
A(N)-\gamma b(N)+\zeta))\]
\newline
Thus the structure function depends only on the function $\rho$.
This function can be arbitrarily chosen and does not influence the
spectrum. In Case 1 we choose
\newline
\begin{equation}
\rho(N) = \frac{1}{3*2^{12}\beta^{8}(N+u)(1+N+u)(1+2(N+u))^{2}}
\end{equation}
\newline
From our expressions for A(N) , b(N) and $\rho(N)$, the third
relation of the cubic associative algebra and the expression of the
Casimir operator we find the structure function $\Phi(N)$. For the
Case 1 the structure function is a polynomial of order 10 in N. The
coefficients of this polynomial are functions of $\alpha$, $\beta$,
$\mu$, $\gamma$, $\delta$, $\epsilon$, $\nu$, $\xi$ and $\zeta$.
\newline
\newline
\textbf{Case 2} $\beta=0$ et $\delta \neq 0$
\begin{equation}
A(N)= \sqrt{\delta}(N+u), b(N)= -\alpha(N+u)^{2} -
\frac{\gamma}{\sqrt{\delta}}(N+u) - \frac{\epsilon}{\delta}
\end{equation}
\newline
In Case 2 we choose a trivial expression $\rho(N)=1$. The explicit
expression of the structure function for this case is
\newline
\begin{equation}
\Phi(N) = (\frac{K}{-4\delta}-\frac{\gamma
\epsilon}{4\delta^{\frac{3}{2}}}-\frac{\zeta}{4\sqrt{\delta}} +
\frac{\epsilon^{2}}{4\delta^{2}})
\end{equation}
\[+(\frac{-\alpha
\epsilon}{2\delta}-\frac{d}{4}-\frac{\gamma^{2}}{4\delta}+\frac{\gamma
\epsilon}{2 \delta^{\frac{3}{2}}}+\frac{\alpha
\gamma}{4\sqrt{\delta}}+\frac{\zeta}{2\sqrt{\delta}}+\frac{\nu\sqrt{\delta}}{12})(N+u)\]
\[+(\frac{-\nu\sqrt{\delta}}{4}-\frac{3\alpha
\gamma}{4\sqrt{\delta}}+\frac{\gamma^{2}}{4\delta}+\frac{\epsilon
\alpha}{2\delta}+\frac{\alpha^{2}}{4}+\frac{\xi}{4}+\frac{\mu\delta}{8})(N+u)^{2}\]
\[+(\frac{-\alpha^{2}}{2}+\frac{\gamma
\alpha}{2\delta^{\frac{1}{2}}}+\frac{\nu\sqrt{\delta}}{6}-\frac{\mu
\delta}{4})(N+u)^{3}+(\frac{\alpha^{2}}{4}+\frac{\mu\delta}{8})(N+u)^{4}\]
\newline
We will consider a representation of the cubic associative algebra
in which the generator A and the Casimir operator K are diagonal. We
use a parafermionic realization in which the parafermionic number
operator N and the Casimir operator K are diagonal. The basis of
this representation is the Fock basis for the parafermionic
oscillator. The vector $|k,n>,n=0,1,2...$ satisfies the following
relations:
\newline
\begin{equation}
N|k,n>=n|k,n>, \quad K|k,n>=k|k,n>
\end{equation}
The vectors $|k,n>$ are also eigenvectors of the generator A.
\[ A|k,n>=A(k,n)|k,n> \]
\begin{equation}
A(k,n)=\frac{\beta}{2}((n+u)^{2}-\frac{1}{4}-\frac{\delta}{\beta^{2}})
 , \beta \neq 0
\end{equation}
\[ A(k,n) = \sqrt{\delta}(n+u) , \beta = 0, \delta \neq 0 \]
\newline
We have the following constraints for the structure function,
\begin{equation}
\Phi(0,u,k)=0 , \quad \Phi(p+1,u,k)=0
\end{equation}
\newline
With these two relations we can find the energy spectrum. Many
solutions for the system exist. Unitary representations of the
deformed parafermionic oscillator obey the following constraint
$\Phi(x)
> 0 $ for x=1,2,...,p .
\newline
\section{Examples}
There exist 21 quantum potentials separable in cartesian coordinates
with a third order integral, we will consider one interesting case
in which the cubic algebra allows us to calculate the energy
spectrum.
\newline
\newline
\textbf{Case Q5}
\newline
\begin{equation}
H = \frac{P_{x}^{2}}{2} + \frac{P_{y}^{2}}{2} + \hbar^{2}(
\frac{x^{2}+y^{2}}{8a^{4}} +
\frac{1}{(x-a)^{2}}+\frac{1}{(x+a)^{2}})
\end{equation}
\begin{equation}
A = \frac{P_{x}^{2}}{2} - \frac{P_{y}^{2}}{2} + \hbar^{2}(
\frac{x^{2}-y^{2}}{8a^{4}} +
\frac{1}{(x-a)^{2}}+\frac{1}{(x+a)^{2}})
\end{equation}
\begin{equation} B = X_{2}= \{L,P_{x}^{2}\} + \hbar^{2}\{y(
\frac{4a^{2}-x^{2}}{4a^{4}} -
\frac{6(x^{2}+a^{2})}{(x^{2}-a^{2})^{2})}),P_{x}\}
\end{equation}
\begin{equation*}
 + \hbar^{2}\{x(\frac{(x^{2}-4a^{2})}{4a^{4}} -
\frac{2}{x^{2}-a^{2}} + \frac{4(x^{2}+a^{2})}{(x^{2}-a^{2})^{2}}
),P_{y}\}
\end{equation*}
\newline
The integrals A,B and H give rise to the algebra
\newline
\[[A,B]=C \]
\begin{equation}
 [A,C]=\frac{h^{4}}{a^{4}}B
\end{equation}
\[ [B,C]= -32\hbar^{2}A^{3} - 48\hbar^{2}A^{2}H + 16\hbar^{2}H^{3}
 + 48\frac{\hbar^{4}}{a^{2}}A^{2} + 32\frac{\hbar^{4}}{a^{2}}HA -
16\frac{\hbar^{4}}{a^{2}}H^{2}\]
\[ + 8\frac{\hbar^{6}}{a^{4}}A - 4\frac{\hbar^{6}}{a^{4}}H -
12\frac{\hbar^{8}}{a^{6}}\quad . \]
\newline
The Casimir operator is
\newline
\begin{equation}
 K = -16\hbar^{2}H^{4} + 32\frac{\hbar^{4}}{a^{2}}H^{3} +
16\frac{\hbar^{6}}{a^{4}}H^{2} - 40\frac{\hbar^{8}}{a^{6}}H -
3\frac{\hbar^{10}}{a^{8}}\quad .
\end{equation}
\newline
and we have
\newline
\begin{equation}
\Phi(x)=(4\frac{a^{4}}{\hbar^{2}}H^{4}-12a^{2}H^{3}+11\frac{\hbar^{4}}{a^{2}}H-\frac{15}{4}\frac{h^{6}}{a^{4}})
\end{equation}
\[+(8a^{2}H^{3}-8\hbar^{2}H^{2}-14\frac{\hbar^{4}}{a^{2}}H-4\frac{\hbar^{6}}{a^{4}})(x+u)+(20\frac{\hbar^{4}}{a^{2}}H-14\frac{\hbar^{6}}{a^{4}})(x+u)^{2}\]
\[+(-8\frac{\hbar^{4}}{a^{2}}H+16\frac{\hbar^{6}}{a^{4}})(x+u)^{3}-4\frac{\hbar^{8}}{a^{4}}(x+u)^{4}\]
\newline
\begin{equation}
 \Phi(x)=(\frac{-4 \hbar^{8}}{a^{4}})(x+u -
(\frac{-a^{2}E}{\hbar^{2}}-\frac{1}{2}))(x+u -
(\frac{a^{2}E}{\hbar^{2}}+\frac{1}{2}))(x+u -
(\frac{-a^{2}E}{\hbar^{2}}+\frac{3}{2}))(x+u -
(\frac{-a^{2}E}{\hbar^{2}}+\frac{5}{2}))
\end{equation}
\newline
We find u with
\[ \Phi(0,u,k)=0 \]
\newline
\begin{equation}
\end{equation}
\[ u_{1}=\frac{-a^{2}E}{\hbar^{2}}-\frac{1}{2},\quad u_{2}=\frac{a^{2}E}{\hbar^{2}}+\frac{1}{2},\quad u_{3}=\frac{-a^{2}E}{\hbar^{2}}+\frac{3}{2},\quad u_{4}=\frac{-a^{2}E}{\hbar^{2}}+\frac{5}{2} \]
\newline
We have four cases:
\newline
Case 1  $u=u_{1}$
\newline
\begin{equation}
E = \frac{\hbar^{2}p}{2a^{2}},\quad
\Phi(x)=(\frac{4\hbar^{8}}{a^{4}})x(p+1-x)(x-2)(x-3)\quad .
\end{equation}
\newline
Case 2 $u=u_{2}$
\begin{equation}
E=\frac{-\hbar^{2}(p+2)}{2a^{2}},\quad
\Phi(x)=(\frac{4\hbar^{8}}{a^{4}})x(p+1-x)(p+3-x)(p+4-x)
\end{equation}
\newline
\begin{equation}
E=\frac{-\hbar^{2}p}{2a^{2}},\quad
\Phi(x)=(\frac{4\hbar^{8}}{a^{4}})x(p+1-x)(p-1-x)(p-2-x)
\end{equation}
\newline
\begin{equation}
E=\frac{-\hbar^{2}(p-1)}{2a^{2}},\quad
\Phi(x)=(\frac{4\hbar^{8}}{a^{4}})x(p+1-x)(p-2-x)(p-x)\quad .
\end{equation}
\newline
Case 3 $u=u_{3}$
\newline
\begin{equation}
E=\frac{\hbar^{2}(p+2)}{2a^{2}},\quad
\Phi(x)=(\frac{4\hbar^{8}}{a^{4}})x(p+1-x)(x-1)(x+2)\quad .
\end{equation}
\newline
Case 4 $u=u_{4}$
\newline
\begin{equation}
E=\frac{\hbar^{2}(p+3)}{2a^{2}},\quad
\Phi(x)=(\frac{4\hbar^{8}}{a^{4}})x(p+1-x)(x+1)(x+3)\quad .
\end{equation}
\newline
The only case that correspond to unitary representations are (3.10)
and (3.14).
\section{Conclusion}
The main results of this article are that we have constructed the
associative algebras for superintegrable potential with a second
order integral and a third order integral. We find realizations in
terms of deformed oscillator algebras for the cubic associative
algebras. We apply our result to a speficic potential$^{13,14}$. We
leave the other quantum cases to a future article.
\newline
We see that many systems in classical and quantum physics are
described by a nonlinear symmetry that provides information about
the energy spectrum.
\newline
We note that our polynomial algebras and their realizations are
independant of the choice of coordinate system. We could apply our
results in the future to systems with a third order integral that
are separable in polar, elliptic or parabolic coordinates. The
method is independant of the metric and we could apply our
polynomial algebras to other cases than superintegrable potentials
in E(2).
\newline
\newline
\textbf{Acknowledgments} The author thanks Pavel Winternitz for his
very helpful comments and suggestions, and also Frederick Tremblay
for useful discussions. The author benefits from FQRNT fellowship.
\section*{\textbf{References}}
$^{1}$V. Bargmann,Zur Theorie des Wasserstoffsatoms, Z. Phys. 99,
576-582 (1936).
\newline
\newline
$^{2}$D.Bonatsos, C.Daskaloyannis and K.Kokkotas, Deformed
oscillator algebras for two-dimensional quantum superintegrable
systems. Phys. Rev.A 50, 3700-3709 (1994).
\newline
\newline
$^{3}$D.Bonatsos, C.Daskaloyannis and K.Kokkotas, Quantum-algebraic
description of quantum superintegrable systems in two dimensions.
Phys. Rev. A 48, R3407-R3410 (1993).
\newline
\newline
$^{4}$C.Daskaloyannis, Quadratic poisson algebras of two-dimensional
classical superintegrable systems and quadratic associative algebras
of quantum superintegrable systems, J.Math.Phys.42, 1100-1119
(2001).
\newline
\newline
$^{5}$C.Daskaloyannis, Generalized deformed oscillator and nonlinear
algebras, J.Phys.A: Math.Gen 24, L789-L794 (1991).
\newline
\newline
$^{6}$J. Drach Sur l'int\'egration logique des \'equations de la
dynamique \`a deux variables: Forces conservatrices. Int\'egrales
cubiques. Mouvements dans le plan, C.R. acad. Sci III, 200, 22-26
(1935).
\newline
\newline
$^{7}$J. Drach, Sur l'int\'egration logique et sur la transformation
des \'equations de la dynamique \`a deux variables: Forces
conservatrices. Int\'egrales. C.R.Acad.Sci III, 599-602 (1935).
\newline
\newline
$^{8}$N.W. Evans. Group theory of the Smorodinsky-Winternitz system.
J. Math. Phys. 32, 3369-3375 (1991).
\newline
\newline
$^{9}$V. Fock, Zur Theorie des Wasserstoffsatoms, Z. Phys.98,
145-154 (1935).
\newline
\newline
$^{10}$J. Fris, V. Mandrosov, Ya.A. Smorodinsky, M. Uhlir and P.
Winternitz, On higher symmetries in quantum mechanics, Phys. Lett.
16, 354-356 (1965).
\newline
\newline
$^{11}$Ya. I Granovskii, A.S. Zhedanov  and I.M. Lutzenko ,
Quadratic Algebra as a Hidden Symmetry of the Hartmann Potential, J.
Phys. A24, 3887-3894 (1991).
\newline
\newline
$^{12}$S.Gravel and P.Winternitz, Superintegrability with
third-order integrals in quantum and classical mechanics J. Math.
Phys. 43, 5902-5912 (2002).
\newline
\newline
$^{13}$S.Gravel, Hamiltonians separable in Cartesian coordinates and
thirdorder integrals of motion, J. Math. Phys. 45, 1003-1019 (2004).
\newline
\newline
$^{14}$S.Gravel, Superintegrability, isochronicity, and quantum
harmonic behavior, ArXiv:math-ph/0310004 (2004).
\newline
\newline
$^{15}$J.M Jauch and E.L Hill, On the problem of degeneracy in
quantum mechanics, Phys Rev, 57, 641-645 (1940).
\newline
\newline
$^{16}$P.Létourneau and L.Vinet, Superintegrable systems: Polynomial
algebras and quasi-exactly solvable hamiltonians, Ann. Physics 243,
144-168 (1995).
\newline
\newline
$^{17}$I.Marquette and P.Winternitz, Polynomial Poissons algebras
for superintegrable systems with a third order integral of motion,
ArXiv:math-ph/0608021 (2006).
\newline
\newline
$^{18}$P. Winternitz, Ya.A Smorodinsky, M.Uhlir et I.Fris, Symmetry
groups in classical and quantum mechanics, Yad.Fiz. 4, 625-635
(1966). (English translation : Sov. J. Nucl. Phys. 4, 444-450
(1967))

\end{flushleft}
\end{document}